\documentclass[amsmath,amssymb,superscriptaddress,showkeys,prd,aps,twocolumn,10pt]{revtex4}
\usepackage{amsmath,color}
\usepackage{graphicx, hyperref}
\usepackage{psfrag}
\usepackage{textcomp}

\newcommand{\R}{R\'enyi }

\RequirePackage{psfrag}
\RequirePackage{textcomp}
\RequirePackage{amsmath,amssymb,color}

\usepackage{ulem}

\begin{document}

\title{Hawking-R\'enyi thermodynamics of rotating black holes from locally Kiselev-type behavior}
%
%

\author{Viktor G.~Czinner}
\email{vczinner(at)gmail.com}
\affiliation
{Pa\c{c}o de Arcos,  Portugal}
\author{Hideo Iguchi}
\email{iguchi.hideo(at)nihon-u.ac.jp}
\affiliation
{Laboratory of Physics, College of Science and Technology, Nihon University,
274-8501 Narashinodai, Funabashi, Chiba, Japan}

\date{\today}


\begin{abstract}
The Hawking-\R model requires the \R entropy thermodynamic temperature of a black hole to be identical with the surface gravity defined-, Hawking temperature. We investigate this approach for stationary black hole space-times, and show that a locally Kiselev-type behavior around the horizon with a coupled anisotropic fluid is sufficient to provide a solution to the problem. In addition, due to the rotating motion, an extra shift has also to be present in the effective mass of the black hole, which is determined by the rotation parameter, $a$, and the \R parameter, $\lambda$. We consider space-times with and without electric charge, and show that the functional form of the solution is the same for both cases. A full thermodynamic analysis of the model falls beyond the scope of this Letter, the main achievements are the derivation and interpretation of the solution.
\end{abstract}

\keywords{black hole thermodynamics, \R entropy, Hawking temperature, stationary space-times, locally Kiselev-type metric, anisotropic fluid}

\maketitle

{\it Introduction}---The \R entropy model to black hole thermodynamics has been around for more than a decade. It was first introduced in \cite{BC,CI,CI2}, and later on many works considered it for various problems in black hole physics \cite{promsiri2020thermodynamics,tannukij2020thermodynamics,nakarachinda2021effective,promsiri2021solid,abreu2021nature,barzi2022renyi,el2022critical,promsiri2022emergent,wang2023thermodynamics,barzi2023some,tong2024topology} and cosmological applications \cite{CM,cc1,cc2,cc3}. In its original form, the model applies the parametric \R entropy formula for black holes, defined as
\begin{equation}\label{SR}
 S_R = \frac{1}{\lambda}\ln(1+\lambda S_{BH})\ ,
\end{equation}
instead of the standard Bekenstein-Hawking entropy \cite{B,Betal,H,H2}, which, for a rotating black hole is given by
\begin{equation}\label{S_BH}
S_{BH}=\pi (r_+^2+a^2)\ .
\end{equation}
The general idea behind the model is to consider the nonextensive nature of strongly gravitating systems in a parametric way. It assumes that black holes may be better described by a more general statistical entropy than the Boltzmann-Gibbs one, such as the Tsallis approach \cite{Tsallis2}, however, in order to have a {\it 0th law} compatible, well defined temperature function, according to the results of general nonadditive thermodynamics \cite{Abe,BV},
the \R entropy \cite{Renyi,Renyibook}, being the {\it formal logarithm} of the Tsallis formula, is applied. For more details on the foundations, please refer to \cite{BC,CI,CI2}.

In the \R entropy approach, the thermodynamic behaviour of black holes is very similar to the original one on anti-de Sitter backgrounds \cite{HP}, and the $\lambda$ parameter of the model can be connected to the cosmological constant, $\Lambda$. The stability properties of the two descriptions were also shown to be identical, exhibiting critical phenomena like Hawking--Page phase transition \cite{HP}, therefore the \R entropy approach can result thermodynamically stable, asymptotically flat black holes. An issue with the \R model however \cite{Nojiri:2021czz}, is that the Hawking temperature, which is defined geometrically via the surface gravity, is in general, different from the \R temperature (derived via the {\it 1st law} of thermodynamics), and only the \R temperature model exhibits the AdS similarity.

Addressing this issue, in a recent paper \cite{HRstatic}, we investigated the question of a possible  equivalence between the two temperatures. We named this approach {\it Hawk\-ing-\R model} (HR for short), based on the requirement that black holes with this property must have their \R temperature, $T_R$, identical with their Hawking temperature, $T_H$. In our work we answered the question of what modification to static black hole metrics can result in the satisfaction of the HR condition. We looked for the solution in the form of a simple metric modifying ansatz, and showed, that for every static, spherically symmetric, vacuum black hole space-time, a corresponding black hole solution can be derived which satisfies the HR criteria. These black holes must be sur\-round\-ed by a certain coupled, anisotropic fluid matter in the form of a Kiselev metric \cite{Kiselev}, where the properties of the fluid are uniquely determined by the mass of the black hole, $M$, and the \R parameter, $\lambda$. We also showed that in the simplest Schwarz\-schild case, the system is thermodynamically unstable, and the {\it 3rd law} of thermodynamics seems to play the role of a cosmic censor.

In this Letter, we are looking at the more general problem of rotating black holes within the HR model. First we consider the Kerr space-time, and by applying the same method we used in the static problem \cite{HRstatic}, we derive the analogous Kiselev-type rotating black hole solution that satisfies the HR condition. We show that in addition to the  anisotropic fluid of the static result, an extra shift must also be present in the effective mass of the black hole, as a consequence of the rotating motion. Furthermore, beyond the rotational effects, we also present a new and very important aspect of the model by pointing out that the Kiselev-type behavior has to be satisfied only locally around the horizon region, and the asymptotic behavior of a general solution can be almost arbitrary, as long as the Einstein equations are globally satisfied.

Due to the complexity of the obtained multi-parametric solution, within this Letter, we cannot look into the details of the thermodynamic and stability properties beyond the fact that it satisfies the Hawk\-ing-\R condition. We leave these investigations to a forthcoming work. Instead, we also consider the problem of an electrically charged, rotating Kerr-Newman black hole, and show that the presence of the electric charge does not alter the functional form of the general solution. Both black holes must have the same modification in their metric tensor in order to result in a corresponding locally Kiselev-type metric that satisfies the HR condition. Throughout this paper, we work in dimensionless units such that $c = G = \hbar = k_B =1$, and the \R entropy parameter, $\lambda$, is also a dimensionless, positive constant.

{\it Kerr space-time}---The Kerr metric in Boyer-Lindquist coordinates is given by
\begin{eqnarray}\label{kerr_metric}
 ds^2 = &&\!\!\!\!\! -\left(1 - \tfrac{2 M r}{\Sigma} \right) dt^2  - \tfrac{4 M r a \sin^2 \theta}{\Sigma} dt d\phi+ \tfrac{\Sigma}{\Delta} dr^2   \nonumber
 \\  &&\!\!\!\!\! + \Sigma d\theta^2 + \left( r^2 + a^2 + \tfrac{2 M r a^2  \sin^2 \theta}{\Sigma}  \right) \sin^2 \theta d \phi^2,
\end{eqnarray}
where
\begin{equation*}
 \Sigma =  r^2 + a^2 \cos^2 \theta , ~~~ \Delta = r^2 + a^2 - 2Mr\ .
\end{equation*}
It is asymptotically flat, $M$ is the space-time's ADM mass and $J\equiv aM$ is the angular momentum, so the rotation parameter, $a$, is the ratio of angular momentum to mass. The Kerr space-time is stationary and axially symmetric, therefore it admits the Killing vectors
\begin{equation}
 t^{\alpha} = \frac{\partial x^{\alpha}}{\partial t} , ~~~
 \phi^{\alpha} = \frac{\partial x^{\alpha}}{\partial \phi}\ ,
\end{equation}
and a null Killing vector, $\xi^{\alpha}$, on the horizon, which is also tangent to the horizon's null generators, can be constructed as
\begin{equation}
 \xi^{\alpha} = t^{\alpha} + \Omega_H\phi^{\alpha},
 \end{equation}
 with
 \begin{equation}\label{Omega}
 \Omega_H \equiv - \left. \frac{g_{t\phi}}{g_{\phi\phi}} \right|_{r_{+}} = \frac{a}{r_{+}^2 + a^2}\ ,
\end{equation}
where $\Omega_H$ is the angular velocity at the outer horizon, $r_{+}$. The surface gravity, $\kappa$, which is independent of $\theta$, can be generally defined as
\begin{equation}\label{kappa}
 \left(-\xi^{\beta}\xi_{\beta}\right)_{;\alpha} = 2\kappa\,\xi_{\alpha}\ , ~~~ \mbox{and} ~~~
 \kappa_{Kerr} = \frac{r_+ - M}{r_+^2 + a^2}\ .
 \end{equation}
Since the Boyer-Lindquist coordinates do not behave well on the horizon, in order to calculate $\kappa$, a coordinate transformation is needed. General techniques are presented e.g.~in \cite{Visser_univ, kappa}, where it is also shown that it can be derived as
\begin{equation}\label{general_kappa}
 \kappa = \frac{(\partial_r\Delta)_{r_+}}{2(r_+^2 + a^2)}\ ,
\end{equation}
very generally for black hole space-times including asymptotically dS/AdS backgrounds as well.

{\it Modified metric ansatz}---The specific question we want to answer is whether there exist a $\lambda$-parametric modified version of the metric (\ref{kerr_metric}), for which the thermodynamic \R temperature of the black hole is identical with the Hawking temperature on the horizon. We search for the solution by requiring a consistent asymptotic behavior: for $\lambda\rightarrow 0$ the Kerr metric should be recovered. In order to satisfy the HR condition, clearly,
the modifying term has to be energy dependent, i.e.~it has to depend on the mass of the black hole. Hence, similarly to our method in solving the static problem \cite{HRstatic}, we start looking for the modified metric with the transformation
\begin{equation}\label{h_trafo}
 M \longmapsto M\left(1+\lambda h(r,a)\right)\ ,
\end{equation}
where $h(r,a)$ may depend on the radial coordinate and the rotation parameter only, but is independent of $\lambda$.
Let us now rewrite the metric (\ref{kerr_metric}) in the following way:
\begin{equation}
 Mr \longmapsto Mg(\lambda,r,a)\ ,
\end{equation}
with
\begin{equation}\label{g_def}
 g(\lambda,r,a) \equiv r\left(1+\lambda h(r,a)\right),
\end{equation}
everywhere, where the term $Mr$ appears. (It should be noted that there is an $r^2$ factor difference between this definition of $g$ and the corresponding definition in \cite{HRstatic}. This, however, shall not result in any confusion of what follows). After this transformation, the $\Delta_g$ component of the modified metric for which we require $T_H\equiv T_R$ becomes
\begin{equation}\label{Delta_g}
 \Delta_g = r^2 + a^2 - 2Mg(\lambda,r,a)\ .
\end{equation}
From the horizon condition, $\Delta_g = 0$, we have
\begin{equation}\label{r_+}
 r_{+} = \sqrt{2Mg(r_+) - a^2}\ ,
\end{equation}
where, and from now on, we explicitly omit indicating the $\lambda$ and $a$ dependence of $g(\lambda, r, a)$, and will only indicate its $r$-dependence when it is taken at the horizon as $g(r_+)$. For the mass of the black hole we have
\begin{equation}\label{M}
 M = \frac{r_+^2 + a^2}{2g(r_+)}\ ,
\end{equation}
and according to (\ref{general_kappa}), the surface gravity of the modified metric can be obtained as
\begin{equation}\label{kappa_g}
\kappa_g = \frac{r_+ - M(\partial_r g)_{r_+}}{r_+^2 + a^2}\ .
\end{equation}

{\it The Hawking-\R problem}---In the \R approach, the {\it 1st law} of thermodynamics is given by
\begin{equation}\label{1stlaw}
 dM = T_R dS_R + \Omega_H dJ\ ,
\end{equation}
and for the HR model we want the Hawking temperature, $T_H$, and the \R temperature, $T_R$, defined respectively as
\begin{equation}\label{temeratures}
T_H=\frac{\kappa_g}{2\pi}\ ,~~~~~
T_R = \left(\frac{\partial M}{\partial S_R}\right)_{\!\!\!J} \ ,
\end{equation}
to be identical. The rotating black hole, as a thermodynamic system, can be described via the extensive quantities as functions of the horizon radius and the rotation parameter as
\begin{equation}
M = M(r_+,a)\ , ~~~ J = J(r_+,a)\ , ~~~ S_R = S_R(r_+,a)\ ,
\end{equation}
where $M$ is given in (\ref{M}), $J=aM$, $S_R$ is defined via (\ref{SR}) and (\ref{S_BH}), and the total differentials are
\begin{eqnarray}
 &&dM = \left(\tfrac{\partial M}{\partial r_+}\right)_{\!\!a} dr_+\
 +\ \left(\tfrac{\partial M}{\partial a}\right)_{\!r_+} da\ ,\\
 &&dJ = \left(\tfrac{\partial J}{\partial r_+}\right)_{\!\!a} dr_+\
 +\ \left(\tfrac{\partial J}{\partial a}\right)_{\!r_+} da\label{dJ}\ ,\\
 &&dS_R = \left(\tfrac{\partial S_R}{\partial r_+}\right)_{\!\!a} dr_+\
 +\ \left(\tfrac{\partial S_R}{\partial a}\right)_{\!r_+} da\ .
\end{eqnarray}
The \R temperature can be derived by keeping the angular momentum constant in the {\it 1st law}, i.e.~$dJ=0$ in (\ref{1stlaw}) as well as in (\ref{dJ}), and the latter results the condition
\begin{equation}\label{da}
 da\ =\ - \left(\frac{\partial J}{\partial r_+}\right)_{\!\!a}\left(\frac{\partial J}{\partial a}\right)^{-1}_{\!r_+}\ dr_+\ ,
\end{equation}
from which we have
\begin{equation}\label{Tg}
T_R\ \equiv\ \left(\frac{\partial M}{\partial S_R}\right)_{\!\!\!J}\ =\
\frac{\left(\frac{\partial M}{\partial r_+}\right)_{\!\!a}-\left(\frac{\partial M}{\partial a}\right)_{\!r_+}\frac{\left(\frac{\partial J}{\partial r_+}\right)_a}{\left(\frac{\partial J}{\partial a}\right)_{r_+}}}
{\left(\frac{\partial S_R}{\partial r_+}\right)_{\!\!a}-\left(\frac{\partial S_R}{\partial a}\right)_{\!r_+}\frac{\left(\frac{\partial J}{\partial r_+}\right)_a}{\left(\frac{\partial J}{\partial a}\right)_{r_+}}}\ .
\end{equation}
Inserting the explicit expressions of $M$, $J $ and $S_R$ results
\begin{eqnarray}\label{TRg}
 T_R = &&\left(2r_+g(r_+) - (r_+^2 + a^2)\tfrac{\partial g(r_+)}{\partial r_+}\right)\times\\
 &&\frac{1+\lambda\pi(r_+^2 + a^2)}{4\pi g(r_+) \left(r_+g(r_+) + a^2\tfrac{\partial g(r_+)}{\partial r_+}
 - ar_+\tfrac{\partial g(r_+)}{\partial a}\right)}\ .\nonumber
\end{eqnarray}
The Hawking temperature, on the other hand, can be simply obtained by plugging (\ref{M}) into (\ref{kappa_g}), and reads as
\begin{equation}\label{THg}
T_H\ \equiv\ \frac{\kappa_g}{2\pi}\ =\
\frac{2r_+g(r_+) - (r_+^2 + a^2)(\partial_r g)_{r_+}}{4\pi g(r_+)(r_+^2 + a^2)}\ .
\end{equation}
By assuming the differential equality
$(\partial_r g)_{r_+}\equiv\tfrac{\partial g(r_+)}{\partial r_+}$
holds for $g$ on the horizon, equating (\ref{TRg}) with (\ref{THg}) results in the following first-order linear partial differential equation,
\begin{eqnarray}\label{gdeq}
 &&a^2\frac{\partial g(r_+)}{\partial r_+} - ar_+\frac{\partial g(r_+)}{\partial a} + r_+g(r_+)\nonumber\\
 &&- (r_+^2 + a^2)\left(1+\lambda\pi(r_+^2 + a^2)\right) = 0\ .
\end{eqnarray}
This PDE can be integrated analytically in a closed form with the general solution
\begin{equation}\label{g}
g(\lambda, r_+, a) = r_+\left[1+\lambda\pi(r_+^2 + a^2)\right]+a\,C(x)\ ,
\end{equation}
where $C(x)$ is an arbitrary differentiable function of the variable $x=\left(r_+^2 + a^2\right)/2$. It is important to notice, that in order to satisfy the HR condition, it is sufficient that the metric deformation function, $g(\lambda, r, a)$, has the form of (\ref{g}) only in an infinitesimal region around the horizon surface, while it can have any different behavior elsewhere, as long as the corresponding metric satisfies the Einstein equations globally, under the conditions imposed so far.
As a consequence, a variety of possibly physically relevant $g$-deformed Kerr black hole metric solutions may exist with a HR thermodynamic behavior. In order to show this, we discuss here the specific case where (\ref{g}) is a global deformation, and find the corresponding exact rotating black hole solution. Other, locally Kiselev-type solutions may be constructed by restricting the global solution to the horizon region as a shell or a domain wall for instance, and matching it with other asymptotic solutions.

{\it Kiselev-type global solution}---Let us consider (\ref{g}) to be globally true, and show that the corresponding metric satisfies the Einstein equations with a rotating black hole interpretation. The integration function, $C(x)$, implies that infinitely many solution exist mathematically, however for a physically realistic one which must be consistent with definition (\ref{g_def}), and can also be referred to as the natural and minimal extension of the corresponding static solution to the stationary case, only the principal part of (\ref{g}) should be considered, so we set $C(x)$  to zero.
As a result, the g-deformed stationary space-time must have the metric function
$h(r,a) = \pi(r^2 + a^2),$
and one can check that transformation (\ref{h_trafo}) indeed provides consistent asymptotic behavior both in the $a\rightarrow 0$ and $\lambda\rightarrow 0$ limits. This $h$ solution is also the exact analogy of what was the result in the static problem, and since $h(r_+,a)=\frac{A}{4}$, with $A$ being the horizon area, the correction can be written in a unified manner as $(1+\lambda A/4)$ on the horizon.

Since the static HR solution resulted in a Kiselev-type black hole metric surrounded by a special, coupled anisotropic fluid \cite{HRstatic}, it is natural to look for a similar interpretation here. For this, it is sufficient to show that the metric function, $\Delta_g$, of the obtained solution
\begin{equation}\label{Delta_gsol}
 \Delta_g = r^2 + a^2 - 2Mr\left[1 + \lambda\pi(r^2 + a^2)\right]\ ,
\end{equation}
is equivalent with a corresponding $\Delta_K$ function,
\begin{equation}\
 \Delta_K = r^2 + a^2 - 2\bar{M}r - Kr^{1-3w}\ ,
\end{equation}
of a rotating Kiselev-type metric, as given e.g.~in \cite{Azreg,Ghosh_rot,Toshmatov,Visser_univ},
where $\bar{M}$ is the mass parameter of the Kiselev black hole, and $K$ and $w$ are constant parameters of a matter field that surrounds it. As it can be easily seen, the equivalence holds simply by setting
\begin{equation}\label{Mbar_K_w}
 \bar M = M(1+\pi\lambda a^2)\ , ~~~K = 2\pi\lambda M\ , ~~~ w = -\frac{2}{3}\ ,
 \end{equation}
so a global rotating black hole solution which satisfies the HR condition can be interpreted as a Kiselev-type black hole with having a shift correction in its effective mass, $\bar M$, compared to the mass, $M$, of the original Kerr black hole. This rotation induced correction is uniquely determined by the parameters $a$ and $\lambda$. The $K$ and $w$ parameters of the matter field are identical to the corresponding ones of the static problem \cite{HRstatic}.

{\it Anisotropic fluid}---As we have shown and thoroughly analyzed it in our previous work, the matter field with the constant parameters, $w = -2/3$ and $K = 2\pi\lambda M$, has an energy-momentum tensor of an anisotropic fluid with
\begin{equation}\label{em_ten_diag}
  T^{ab} =\text{diag}( \rho,  p_r,  p_t,  p_t) \ ,
\end{equation}
where the static energy density, $\rho$, and the static an\-isotropic pressure components, $p_r$ and $p_t$ were
\begin{equation}
 \rho = - p_r = -2 p_t = \frac{\lambda M}{2 r} \ .
\end{equation}
(Unfortunately a minus sign mistake was made in the corresponding equation (20) in \cite{HRstatic}, here we use the corrected form.) In the present, rotating space-time, by using the standard orthogonal tetrad
\begin{eqnarray}
 && e^\mu_{(t)} = \left( \tfrac{r^2 + a^2}{\sqrt{\Delta_g \Sigma}}, 0, 0, \tfrac{a}{\sqrt{\Delta_g \Sigma}} \right),
 ~~e^\mu_{(r)} = \left( 0, \sqrt{\tfrac{\Delta_g}{\Sigma} }, 0, 0 \right),\nonumber\\
 && e^\mu_{(\theta)} = \left( 0, 0, \tfrac{1}{\sqrt{\Sigma}}, 0 \right),
 ~~~~~~e^\mu_{(\phi)} = \left( \tfrac{a \sin \theta}{\sqrt{\Sigma}}, 0, 0, \tfrac{1}{\sin \theta \sqrt{\Sigma}}\right),\nonumber
\end{eqnarray}
the energy-mo\-mentum tensor can also be presented in the diagonalized form of (\ref{em_ten_diag}), and the energy density and pressure components can be derived from the Einstein tensor as
\begin{eqnarray}
 &&\!\!\!\!\!\!\!\!\!\!\!\!\!\! \rho  = \frac{G_{\mu\nu} e^\mu_{(t)}e^\nu_{(t)} }{8 \pi}  = \frac{\lambda M r^{3}}{2 \Sigma^2},~~
 p_r = \frac{G_{\mu\nu} e^\mu_{(r)}e^\nu_{(r)} }{8 \pi} = - \rho,\\
 &&\!\!\!\!\!\!\!\!\!\!\!\!\!\! p_\theta = \frac{G_{\mu\nu} e^\mu_{(\theta)}e^\nu_{(\theta)} }{8 \pi} = - \frac{\lambda M r \left( r^2 + 3 a^2 \cos^2 \theta \right)}{4 \Sigma^2}\ , \\
 &&\!\!\!\!\!\!\!\!\!\!\!\!\!\! p_\phi = \frac{G_{\mu\nu} e^\mu_{(\phi)}e^\nu_{(\phi)} }{8 \pi}  =  p_\theta,
\end{eqnarray}
describing the very same fluid around a rotating black hole with also providing a consistent $a\rightarrow 0$, static limit.

{\it Null-energy condition}---As shown in \cite{HRstatic,vprd}, the corresponding matter, in the static limit, satisfies the NEC, and may be interpreted as a coupled and electrically charged anisotropic fluid which is supported by both pressure gradients and its own internally generated an\-isotropic electric field. For more details please refer to \cite{HRstatic,vcqg,vprd}. In the present rotating case, the NEC also requires $(\rho + p_\theta) \geq 0$, that is
$(r^2 - 3 a^2 \cos^2\theta)\ge 0$,
which, for general $\theta$ simplifies to $r\geq \sqrt{3}a$. For sufficiently slowly rotating black holes, this condition is satisfied globally for $r\ge r_+$, however, for space-times with larger rotation parameters, it may be violated in the near horizon region. In order to obtain the exact value of $a_{max}$, where the NEC is still globally satisfied, a detailed horizon structure analysis is necessary, which goes beyond the limits of this Letter. We intend to present the corresponding results in a forthcoming publication together with the detailed thermodynamic and stability analysis. The important point is that NEC satisfying, physically realistic, rotating solutions do exist with slow enough rotation.

{\it Black holes with electric charge}---For the case of an electrically charged, Kerr-Newman black hole, it is easy to see that the functional form of the corresponding $g$-deformed solution remains the same. The metric function, $\Delta_g$, in this case becomes
\begin{equation}\label{Delta_gq}
 \Delta_g = r^2 + a^2 + Q^2 - 2Mg(\lambda,r,a)\ ,
\end{equation}
but the surface gravity, $\kappa_g$, has the very same functional form of (\ref{kappa_g}), only the black hole mass gets modified as
\begin{equation}\label{M_q}
 M = \frac{r_+^2 + a^2+ Q^2}{2g(r_+)}\ ,
\end{equation}
due to the presence of the electric charge, $Q$, via the horizon condition $\Delta_g=0$. Of course the {\it 1st law} has also to be extended with the electric term as
\begin{equation}\label{1stlaw_q}
dM = T_R dS_R + \Omega_H dJ + \Phi_H dQ\ ,~~~~ \Phi_H = \frac{r_+Q}{r_+^2+a^2}\ ,
\end{equation}
where $\Phi_H$ is the electrostatic potential at the horizon. When computing the \R temperature from
\begin{equation}\label{TRq}
T_R = \left(\frac{\partial M}{\partial S_R}\right)_{\!\!\!J,\,Q} \ ,
\end{equation}
in addition to the angular momentum, the electric charge has to be kept constant as well, i.e.~$dJ=dQ=0$, so the formula obtained in (\ref{Tg}) remains unchanged and results in
\begin{eqnarray}\label{TRgq}
 T_R\ = &&\left(2r_+g(r_+) - (r_+^2 + a^2+ Q^2)\tfrac{\partial g(r_+)}{\partial r_+}\right)\times\\
 &&\frac{1+\lambda\pi(r_+^2 + a^2)}{4\pi g(r_+) \left(r_+g(r_+) + a^2\tfrac{\partial g(r_+)}{\partial r_+}
 - ar_+\tfrac{\partial g(r_+)}{\partial a}\right)}\ .\nonumber
\end{eqnarray}
By equating $T_H$ with $T_R$, due to the cancellation of the first bracket term, we obtain the very same PDE (\ref{gdeq}) for $g$, as in the case of the uncharged $g$-deformed Kerr metric.

Accordingly, the presence of the electric charge does not modify the general solution, $g(\lambda,r_+,a)$, obtained in (\ref{g}). For the Kiselev-type global metric, in addition to the electric field component in the energy-momentum tensor, the very same, coupled anisotropic fluid, together with the same shift in the black hole mass is sufficient to satisfy the HR condition. The total energy-momentum tensor, $T_{\mu\nu}$, of the $g$-deformed Kerr-New\-man metric is the sum of the fluid and electric fields with energy density and pressure components
\begin{eqnarray}
&&\rho  = - p_r =  \frac{\lambda M r^{3}}{2 \Sigma^2} + \frac{Q^2}{8\pi \Sigma^2}\ , \\
&&p_\theta = p_\phi = - \frac{\lambda M r \left( r^2 + 3 a^2 \cos^2 \theta \right)}{4 \Sigma^2} + \frac{Q^2}{8\pi \Sigma^2}\ . \end{eqnarray}

{\it Conclusions}---In this Letter we extended the Hawking-\R model of \cite{HRstatic} in two fundamental ways. First by considering more general, stationary black hole space-times, and second by showing that general solutions to the HR problem do not need to be Kiselev-type globally, it is sufficient to exhibit this property only locally, in the near horizon region. This results that a variety of black hole solutions, possibly even asymptotically flat or cosmological ones, may exist that can satisfy the HR criterion. The HR condition restricts the behavior of the $g$-function on the black hole horizon only, in other regions it can be arbitrary, as long as the corresponding metric satisfies the Einstein equations. By assuming that  (\ref{g}) holds for $g$ even globally, we showed that it describes a rotating Kiselev-type black hole metric with the very same coupled anisotropic fluid that was obtained for the static problem, and for sufficiently slow rotation this fluid satisfies the NEC. In addition to the presence of the fluid matter, the effective mass parameter of the new rotating black hole has also to be increased compared to the mass of the original Kerr metric, and the increment is uniquely determined by the parameters $a$ and $\lambda$.

Unlike in the static case, in this Letter we didn't analyse the horizon structure and the thermodynamic- and stability properties of the solution in details. Instead we also considered the problem of electrically charged rotating black holes, and showed,  that the general condition, (\ref{g}), remains formally the same in the case of a $g$-deformed Kerr-Newman metric, and the same, locally Kiselev-type, anisotropic fluid solution satisfies the HR condition in the presence of electric charge as well. Although we considered asymptotically flat space-times only, one can show that the approach also works on asymptotically dS/AdS backgrounds as well. In conclusion, our findings are relevant in showing the generality of the HR thermodynamic model by extending its validity to a wider spectrum of black hole solutions, including shell- or domain-wall like asymptotically flat or cosmological scenarios as well. We hope to report on our findings of a detailed thermodynamic analysis soon, in a forthcoming publication.



\bibliography{references}

\end{document}